\documentclass[twocolumn,aps,showpacs,prc]{revtex4}
\usepackage[dvips]{epsfig}
\begin{document}

\title{ Sub-barrier fusion hindrance and absence of neutron transfer channels }

\author{ Vinay Singh$^{1}$, Joydev Lahiri$^{1}$, Partha Roy Chowdhury$^{2}$ and D. N. Basu$^{1}$}

\affiliation{$^1$Variable Energy Cyclotron Centre, 1/AF Bidhan Nagar, Kolkata 700064, INDIA}
\affiliation{ $^2$ Chandrakona Vidyasagar Mahavidyalaya, Chandrakona Town, Paschim Medinipur, West Bengal 721201, India}

\email[E-mail 1: ]{vsingh@vecc.gov.in}
\email[E-mail 2: ]{joy@vecc.gov.in}
\email[E-mail 3: ]{royc.partha@gmail.com}
\email[E-mail 4: ]{dnb@vecc.gov.in} 

\date{\today }

\begin{abstract}

    The sub-barrier fusion hindrance has been observed in the domain of very low energies of astrophysical relevance. This phenomenon can be analyzed effectively using an uncomplicated straightforward elegant mathematical formula gleaned presuming diffused barrier with a Gaussian distribution. The mathematical formula for cross section of nuclear fusion reaction has been obtained by folding together a Gaussian function representing the fusion barrier height distribution and the expression for classical cross section of fusion assuming a fixed barrier. The variation of fusion cross section as a function of energy, thus obtained, describes well the existing data on sub-barrier heavy-ion fusion for lighter systems of astrophysical interest. Employing this elegant formula, cross sections of interacting nuclei from $^{16}$O + $^{18}$O to $^{12}$C + $^{198}$Pt, all of which were measured down to $<$ 10 $\mu$b have been analyzed. The agreement of the present analysis with the measured values is comparable, if not better, than those calculated from more sophisticated calculations. The three parameters of this formula varies rather smoothly implying its usage in estimating the excitation function or extrapolating cross sections for pairs of interacting nuclei which are yet to be measured. Possible effects of neutron transfers on the hindrance in heavy-ion fusion have been explored.

\vskip 0.2cm
\noindent
{\it Keywords}: Fusion reactions; Excitation function; Barrier distribution; Nucleosynthesis.  
\end{abstract}

\pacs{ 25.70.-z; 26.20.Np; 25.60.Pj; 97.10.Cv }   
\maketitle

\noindent
\section{Introduction}
\label{section1}

    The phenomenon of fusion hindrance in heavy-ion fusion reactions for lighter systems may have important consequences on the nuclear processes taking place in astrophysical scenarios. The hindrance can seriously affect the energy generation by heavy-ion fusion reactions occurring in the region of extreme sub-barrier energies, encompassing reactions involving lighter systems, such as the reactions occurring in the stage of carbon and oxygen burning in heavy stars \cite{Ro91,Da98,Wa97} and their evolution and elemental abundances. It is well recognized that excitation functions for fusion of two colliding nuclei can not be explained satisfactorily within the framework of a single barrier penetration that is well-defined in its total potential energy. While replicating the shapes of excitation functions for fusion reactions, particularly in the domain of low energies near-threshold, coexistence of different barriers must be assumed. Describing the fusion cross section calculations within the framework of any theory of nuclear reactions which involves coupling to several collective states \cite{We91,Ste95,Ti98,Tr01,St06,St07,Den14,Den19}, this condition is naturally accounted.

    The intent of the current exercise is to estimate the cross sections for nuclear reactions in heavy-ion fusion of lighter systems of astrophysical interest. The phenomenological description of the dependence of nuclear fusion reactions on the collisional kinetic energy is accomplished by assuming the Gaussian form of fusion barrier distribution and considering the mean height of barriers and the variance as freely varying parameters. The cross sections for fusion reactions have been obtained by folding a Gaussian distribution for fusion barrier heights together with the classical expression of fixed single barrier cross section for fusion. The effective radius, which is the distance of the position of the potential barrier of interacting nuclei, is treated as an additional free parameter. These three parameters can be extracted distinctively for each reaction by fitting the theoretical predictions to the experimental data. The excitation functions for fusion reactions, thus obtained, describes well existing data of sub-barrier fusion and energy dependence of capture reactions for lighter heavy-ion systems of astrophysical interest.

\noindent
\section{Distribution of fusion barriers}
\label{section2}

    It has been observed that the excitation functions for fusion of two colliding nuclei can not be well explained by invoking a well-defined single one-dimensional potential barrier penetration model. The explanation of fusion cross sections, particularly in case of heavy-ions, requires invoking of a potential barrier distribution \cite{Ro91}. The quantum mechanical barrier penetration smoothens out a set of distinct potential energy barriers to an effective barrier distribution which is continuous. In order to replicate the dependence of nuclear cross sections of fusion reaction on the collisional kinetic energy, specifically measured at low energies near fusion threshold, the assumption of a fusion barrier height distribution becomes necessary to simulate the effects resulting from the coupling to other channels. In the coupled-channel calculations it is naturally achieved which involve, in both colliding nuclei, the coupling to collective states down to the lowest level. The nuclear structure effects influencing the distribution of potential energy barriers have been considered negligible and hence ignored in this work. A Gaussian form $D(h)$ simulating the shape of the diffused barrier has been conceptualized \cite{Wi04} for the fusion barrier height distribution. Therefore, the distribution of barriers is provided by

\begin{equation}
 D(h)=\frac{1}{\sqrt{2\pi}\sigma_h}\exp\Big[-\frac{(h-h_0)^2}{2\sigma_h^2}\Big] 
\label{seqn1}
\end{equation}
\noindent
where for each individual reaction, the parameters $h_0$ (mean barrier height) and $\sigma_h$ (width of barrier distribution) are to be determined exclusively.

\begin{figure}[t]
\vspace{0.8cm}
\eject\centerline{\epsfig{file=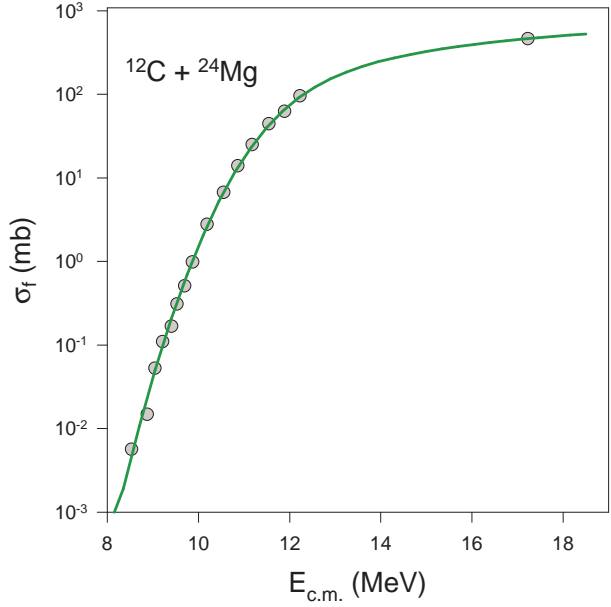,height=8cm,width=8cm}}
\caption
{Plots of the analytical estimates (solid line) and the measured values (full circles) of the capture excitation functions for $^{12}$C+$^{24}$Mg.}
\label{fig1}
\vspace{0.0cm}
\end{figure}
\noindent 

\begin{figure}[t]
\vspace{0.8cm}
\eject\centerline{\epsfig{file=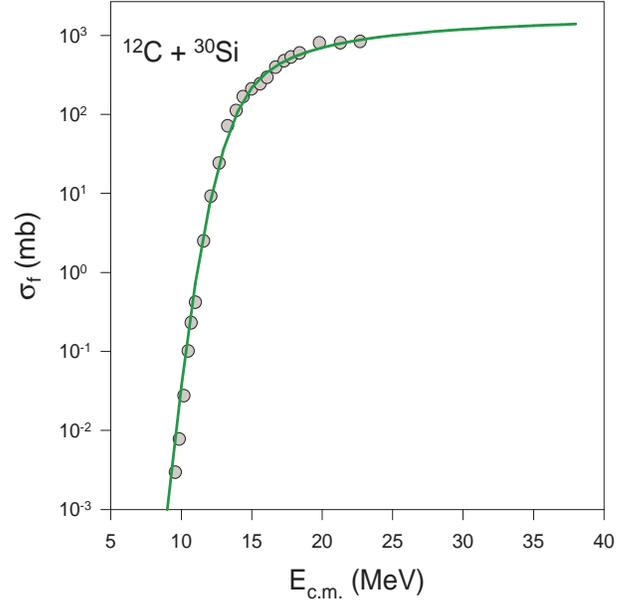,height=8cm,width=8cm}}
\caption
{Plots of the analytical estimates (solid line) and the measured values (full circles) of the capture excitation functions for $^{12}$C+$^{30}$Si.}
\label{fig2}
\vspace{0.0cm}
\end{figure}
\noindent 

\noindent
\section{Fusion cross section calculation}
\label{section3}

    In furtherance of providing a systematic analysis to the excitation function measurements of nuclear fusion reactions, a mathematical formula for the cross section can be derived \cite{Wi04} for surmounting the barrier arising due to interacting nuclei. The energy dependency of the cross sections for nuclear fusion reactions is accomplished by folding the Gaussian distribution for fusion barriers \cite{Wi04,Ca11} given by Eq.(1) together with the classical nuclear fusion reaction cross section expression which is provided by

\begin{eqnarray}
 \sigma_{f}(h) =&& \pi R_h^2 \Big[1-\frac{h}{E}\Big] ~~~~~~~~~~~~~~{\rm for} ~~h\leq E  \nonumber\\
 =&&0 ~~~~~~~~~~~~~~~~~~~~~~~~~~~~~~{\rm for}~~h\geq E
\label{seqn2}
\end{eqnarray}
\noindent
where $R_h$ marks, approximately, the position of the barrier (effective radius corresponding to the relative distance), which results in the following expression
  
\begin{eqnarray}
 &&\sigma_c(E) = \int_{E_0}^\infty \sigma_{f}(h) D(h)dh \nonumber\\
 &&= \int_{E_0}^{h_0} \sigma_{f}(h) D(h)dh + \int_{h_0}^E \sigma_{f}(h) D(h)dh  \\
 &&= \pi R_h^2\frac{\sigma_h}{E\sqrt{2\pi}}\Big[\xi\sqrt{\pi}\Big\{{\rm erf}(\xi)+{\rm erf}(\xi_0)\Big\}   
 +e^{-\xi^2} -e^{-\xi_0^2}\Big]  \nonumber
\label{seqn3}
\end{eqnarray}
\noindent
where $E_0=0$ for positive $Q$ value reactions and $E_0=Q$ for negative $Q$ value reactions, $Q$ value being the sum of the rest masses of fusing nuclei minus rest mass of the resultant fused nucleus, 

\begin{eqnarray}
 \xi &&= \frac{E-h_0}{\sigma_h\sqrt{2}} \nonumber\\
 \xi_0 &&= \frac{h_0-E_0}{\sigma_h\sqrt{2}} 
\label{seqn4}
\end{eqnarray}
\noindent
and erf($\xi$) is the Gaussian error integral for argument $\xi$. The three parameters $R_h$, $h_0$ and $\sigma_h$ have to be determined by a least square fitting of Eq.(3), while making use of Eq.(4), to the measured excitation functions for fusion reactions. While deriving the formula of Eq.(3), the quantal effects of barrier penetration have not been taken explicitly into account. The structure of a given excitation function for fusion reaction is, however, influenced by the sub-barrier tunneling which has been included effectively through the parameter $\sigma_h$ which describes the width of barrier distribution. 

    The mathematical expression of Eq.(3), for the interaction cross section of overcoming the barrier arising due to potential-energy, is achieved by the use of a diffused-barrier. This formula provides a very elegant parametrization for such cross sections. Hence, for the predictions and analysis of excitation functions for fusion reactions, particularly in the span of energies below barrier, it may be used effectively for systems involving light, medium or moderately heavy ions. 

    In case of systems involving light or medium heavy ions, fusion is automatically guaranteed that leads to compound nucleus formation, once it surmounts the barrier of the colliding nuclei. The word `capture' refers to the action of surmounting the potential barrier of colliding ions that follows a composite system formation. In general, there is a probability $f$ that composite nucleus experiences fusion in an event of target nucleus capturing the projectile. This probability approaches unity for light and medium systems. Under this condition ($f\sim1$) fusion ensues for most of the capture events leading to capture cross sections being practically identical to fusion cross sections. On the contrary, there is only a meager probability ($f<1$) that events leading to capture would eventually proceed to fusion in case of very heavy systems. In these cases most of the events remaining re-separate before equilibration. For such cases distinguishing fusion from capture becomes necessary. Therefore, in the event of quite heavy systems, estimations obtained using Eq.(3) will produce capture cross sections where fusion is not automatically guaranteed once the barrier penetration is complete.

\begin{figure}[t]
\vspace{0.8cm}
\eject\centerline{\epsfig{file=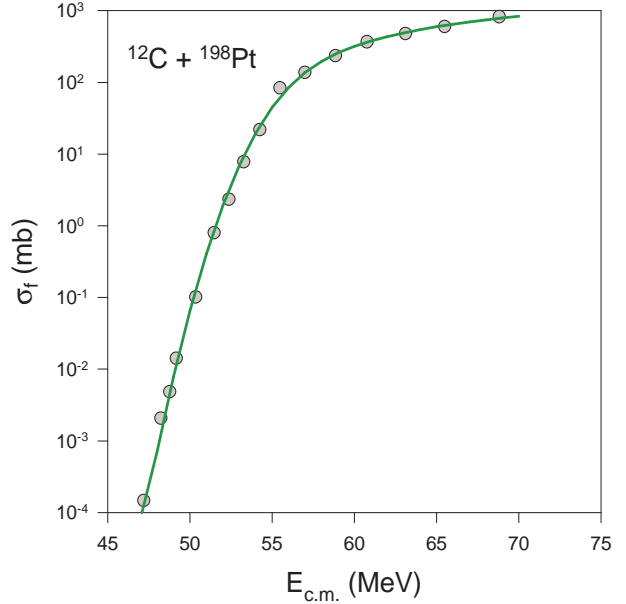,height=8cm,width=8cm}}
\caption
{Plots of the analytical estimates (solid line) and the measured values (full circles) of the capture excitation functions for $^{12}$C+$^{198}$Pt.}
\label{fig3}
\vspace{0.0cm}
\end{figure}
\noindent
  
\begin{figure}[t]
\vspace{0.8cm}
\eject\centerline{\epsfig{file=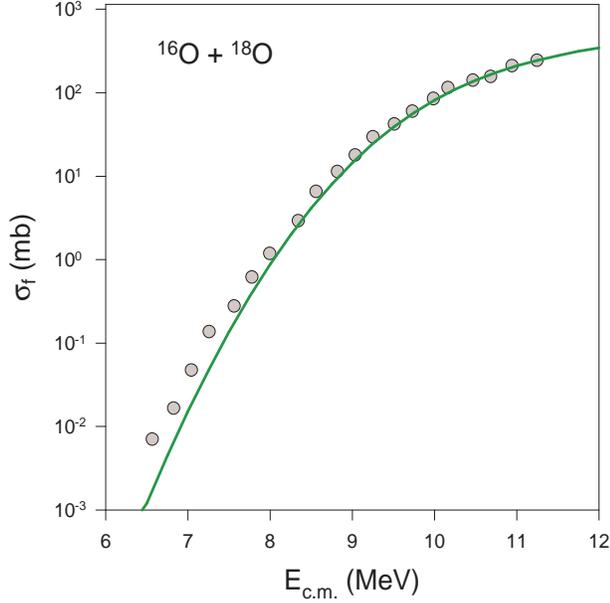,height=8cm,width=8cm}}
\caption
{Plots of the analytical estimates (solid line) and the measured values (full circles) of the capture excitation functions for $^{16}$O+$^{18}$O.}
\label{fig4}
\vspace{0.0cm}
\end{figure}
\noindent
        
\noindent
\section{ Results and discussion }
\label{section4}

\subsection{ The fusion excitation functions }

    The excitation functions for nuclear reactions of heavy-ion fusion for lighter systems of astrophysical interest at sub-barrier energies have been analyzed theoretically. This has been facilitated through Eq.(3) which is obtained by folding an elegant diffused barrier distribution [provided in Eq.(1)] of Gaussian shape together with the classical mathematical fusion cross section expression for single fixed barrier. Identical combinations of projectile-target system involved in heavy ion sub-barrier fusion reactions have been chosen which have been recently \cite{Mo20,Mo97,Sh16,We21,Ra20} studied. By using the method of least-square fitting, the values of three parameters $h_0$ (mean barrier height), $\sigma_h$ (width) and $R_h$ (effective radius) have been extracted. The values of these are tabulated in Table-I arranging it in the ascending order of projectile masses. 

\begin{figure}[t]
\vspace{0.8cm}
\eject\centerline{\epsfig{file=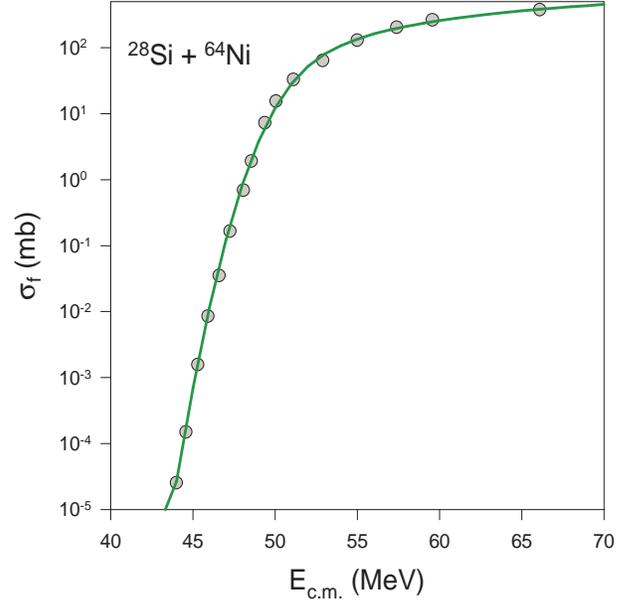,height=8cm,width=8cm}}
\caption
{Plots of the analytical estimates (solid line) and the measured values (full circles) of the capture excitation functions for $^{28}$Si+$^{64}$Ni.}
\label{fig5}
\vspace{0.0cm}
\end{figure}
\noindent 

\begin{figure}[t]
\vspace{0.8cm}
\eject\centerline{\epsfig{file=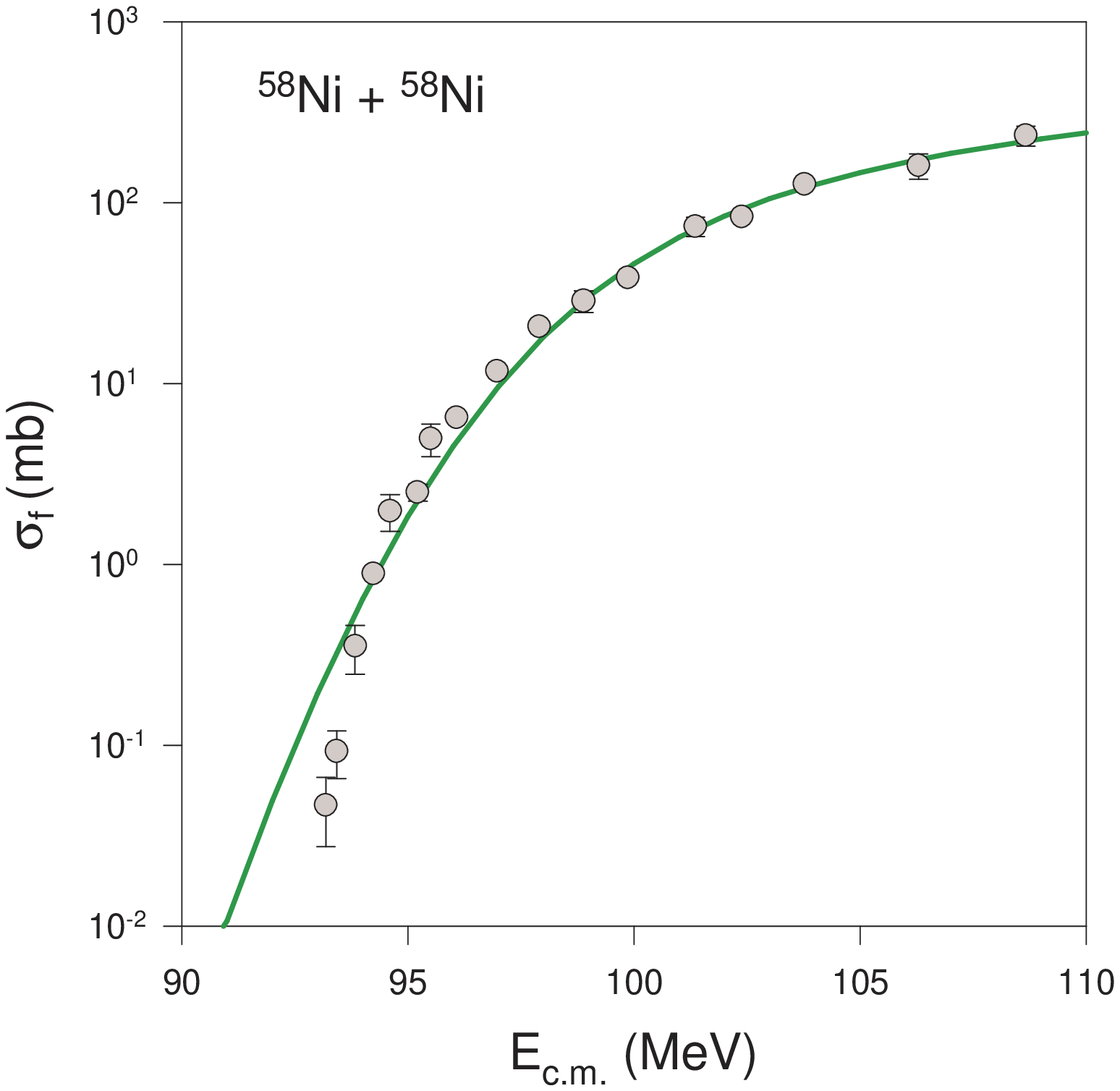,height=8cm,width=8cm}}
\caption
{Plots of the analytical estimates (solid line) and the measured values (full circles) of the capture excitation functions for $^{58}$Ni+$^{58}$Ni.}
\label{fig6}
\vspace{0.0cm}
\end{figure}
\noindent 

    The task of estimating cross section $\sigma_c(E)$ for a particular reaction rests upon predicting the parameter values for $h_0$ (mean barrier height), $\sigma_h$ (width) and $R_h$ (effective radius) reliably. Since $h_0$ is necessarily the mean barrier height, it should be a function of $z = Z_1Z_2/(A_1^{1/3} + A_2^{1/3})$ (Coulomb parameter) in neighborhood of the fusion barrier. The third entity, the effective barrier radius $R_h$, unquestionably should depend upon $r_0(A_1^{1/3} + A_2^{1/3}) = r_0 A_{12}$ where $r_0$ is the nuclear radius parameter. The extrapolation of the tendencies of $\sigma_h$ is extra difficult, which basically arise due to nuclear deformation, vibrations and quantum mechanical barrier tunneling probability. It may be observed from Table-I that the mean barrier height $h_0$ increases with the Coulomb parameter $z$ for all cases while the effective radius $R_h$ also increases with $A_{12}$ except for $^{16}$O+$^{18}$O and $^{28}$Si+$^{64}$Ni and interestingly, as may be seen from Figs.-1-7, for $^{16}$O+$^{18}$O case, the theoretical calculations are slightly off at lower energies as well.      

\begin{table}[ht]
\vspace{0.0cm}
\caption{\label{tab:table1} The extracted values of $h_0$ (mean barrier height), $\sigma_h$ (width) and $R_h$ (effective radius), obtained from the analyses of the measured fusion excitation functions. The table is arranged in order of increasing projectile mass.}
\begin{tabular}{|c|c|c|c|c|c|c|}
\hline
Reaction &Refs.&$z$&$A_{12}$&$\sigma_h$&$h_0$&$R_h$ \\
& &  &  & [MeV] & [MeV] & [fm]         \\  
\hline

$^{12}$C+$^{24}$Mg&  \cite{Mo20}&13.916 &5.174  &0.815 &11.483 &6.646 \\   
$^{12}$C+$^{30}$Si&  \cite{Mo97}&15.565 &5.397  &1.090 &13.540 &8.300 \\  
$^{12}$C+$^{198}$Pt& \cite{Sh16}&57.650 &8.118  &1.749 &55.140 &11.179 \\ 
$^{16}$O+$^{18}$O&   \cite{Mo20}&12.450 &5.141  &0.859 &9.797 &7.743 \\
$^{28}$Si+$^{64}$Ni& \cite{We21}&55.709 &7.037  &1.402 &50.403 &7.182 \\
$^{58}$Ni+$^{58}$Ni& \cite{Ra20}&101.269&7.742  &2.275 &98.278 &8.550 \\ 
$^{64}$Ni+$^{64}$Ni& \cite{Ra20}&98.000 &8.000  &1.466 &92.646 &8.862 \\
\hline
\end{tabular} 
\vspace{0.0cm}
\end{table}

    In Figs.-1-7, the fusion excitation functions which are measured experimentally have been depicted by full circles and compared with the predicted estimates obtained using the diffused fusion barrier formula represented by the continuous lines. Results for the colliding $^{12}$C+$^{24}$Mg, $^{12}$C+$^{30}$Si, $^{12}$C+$^{198}$Pt, $^{16}$O+$^{18}$O, $^{28}$Si+$^{64}$Ni, $^{58}$Ni+$^{58}$Ni and $^{64}$Ni+$^{64}$Ni systems are illustrated in Figs.-1-7. It may be easily identified from the plots that accurately measured excitation functions for fusion reactions yield a systematized information on the cardinal attributes of the nucleus-nucleus interaction potential, {\it viz.} $h_0$ (mean barrier height) and $\sigma_h$ (width) of its distribution for collisions between two nuclei. The capture or fusion cross sections for planning experiments can be also guessed using Eq.(3) along with the theoretically extracted values of $h_0$ and $\sigma_h$ parameters. 
    
    As can be visualized from the Figs.-1-7, the theoretical estimates facilitated by the diffused barrier formula described in this work resulted in good fits to the measured experimental data. This observation obviously infers that almost all the events leading to capture proceed to fusion for the chosen set of nuclei resulting in capture cross sections being essentially identical to the fusion cross sections. It may be further imply that the for the barrier distribution, the choice of Gaussian form describes quite well the nuclear cross sections for fusion reactions at energies below the barrier. This fact justifies the model `beyond single barrier' which arises out of vibration and deformation of nuclei and more importantly tunneling. Whereas theoretically `barrier distribution' is a valid concept under a few approximations, the fact that fits to the experimentally measured data are good implies certainly that in fusion reactions involving heavy-ions, it remains a meaningful concept at least for lighter systems of astrophysical interest. 
    
\begin{figure}[t]
\vspace{0.8cm}
\eject\centerline{\epsfig{file=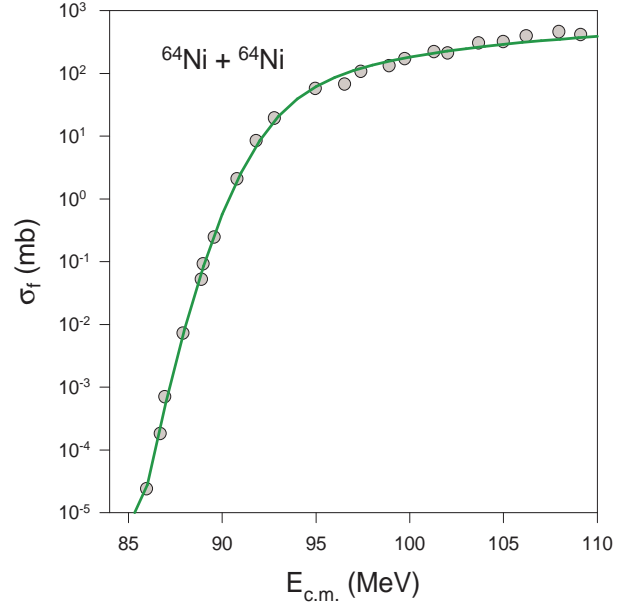,height=8cm,width=8cm}}
\caption
{Plots of the analytical estimates (solid line) and the measured values (full circles) of the capture excitation functions for $^{64}$Ni+$^{64}$Ni.}
\label{fig7}
\vspace{0.0cm}
\end{figure}
\noindent  

    It is pertinent to mention here that although the single-Gaussian parameterization [Eq.(1)] for barrier distribution is reasonably successful in providing a good description of fusion process in general, neither the formula derived for fusion cross-section nor the method using the barrier distribution can be put to use for all fusing systems. One can visualize from Eq.(3) that the excitation function is all the time a monotonically rising function of energy. This puts a limitation on Eqs.(1,2) which can not be used for describing fusion reactions at higher energies when incomplete fusion as well as deep-inelastic scattering can cause a lowering of the fusion cross section. Similar limitation arises for lighter systems (e.g. $^{12}$C+$^{12}$C, $^{12}$C+$^{16}$O, $^{16}$O+$^{16}$O etc.) as well when excitation functions possess oscillations and resonance structures. The possibility of better agreement to data may further be explored by opting a more intricate formula for the barrier distribution, which, nonetheless, will bring in more additional adjustable parameters than just three used in the present work. Such refinements for the barrier distribution of Eq.(1) may include distributions having different widths on lower or higher energy sides, certain moderation of the exponent in Gaussian distribution form or multi-component distributions.

\subsection{ The fusion barrier distribution parameters }
    
    The anticipation of the swing of $\sigma_h$ is easier said than done. The reason lies in the fact that $\sigma_h$ crops up mainly due to nuclear vibrations, deformation and quantum mechanical barrier tunneling probability. A nucleus (tagged $i$) with a static deformation of magnitude $\beta_2(i)$ can have all possible orientations which lead to a standard deviation (SD) of $\Delta R_i$ \cite{Es81} in the effective radius $R_i$ given by

\begin{equation}
 \Delta R_i = \frac{\beta_2(i)R_i}{\sqrt{4\pi}}
\label{seqn6}
\end{equation}
\noindent      
where except for the quadrupole, all other higher multipoles have been neglected. Thus, for a fixed distance between the centers of mass of two nuclei, the distribution of the sub-sequential surface-surface distance results in the SD $\sigma_i$ for the barrier height distribution given by

\begin{eqnarray}
 \sigma_i=&&\bigg|\frac{\partial V}{\partial r}\bigg|_{r=R_h} \Delta R_i \nonumber\\
 =&& \frac{Z_1Z_2e^2}{R_h}  \frac{\beta_2(i)}{\sqrt{4\pi}} \frac{R_i}{R_h} \Big[1 - \frac{3R_i}{5R_h} \Big]
\label{seqn7}
\end{eqnarray}
\noindent  
Therefore, $\sigma_h$ can be given by

\begin{equation}
 \sigma_h = \sqrt{\sigma_1^2 + \sigma_2^2 + \sigma_0^2} 
\label{seqn8}
\end{equation}
\noindent
where $R_h=R_1+R_2=r_0[A_1^{1/3} + A_2^{1/3}]$, $\Delta R_1$ and $\Delta R_2$ are the SDs of the radius vectors specifying the surfaces of the target as well as the projectile nuclei having mean radii $R_1$ and $R_2$ and quadrupole deformation parameters $\beta_2(1)$ and $\beta_2(2)$, respectively. The quantity $\sigma_0$ in the above Eq.(7) is an adjustable parameter which, at least approximately, takes into account the nuclear vibrations and quantum mechanical barrier tunneling probability. Manifestly, for semi-magic as well as magic nuclei, $\sigma_1 = \sigma_2 = 0$ and then $\sigma_h = \sigma_0$.

    The fusion $Q$ values, being the sum of the rest masses of fusing nuclei minus rest mass of the resultant fused nucleus, have been calculated using atomic mass excesses \cite{Au03}. In Table-II, the fusion $Q$ values, the effective radius parameter $r_0$ and the quantities $\sigma_1$, $\sigma_2$ and $\sigma_0$, obtained from the analyses of the measured fusion excitation functions, have been listed. For obtaining the values of $\sigma_1$ and $\sigma_2$, the theoretical values of static deformation $\beta_2(i)$ \cite{Mo95} from recent tabulation \cite{Mo16} have been used while the measured nuclear deformations are available in Ref.\cite{Ra01}.  

\begin{table*}[ht]
\vspace{0.0cm}
\caption{\label{tab:table2} The list of extracted values of effective radius parameter $r_0$ and the quantities $\sigma_1$, $\sigma_2$ and $\sigma_0$, obtained from the analyses of the measured fusion excitation functions is presented. The table is arranged in order of increasing projectile mass along with the fusion $Q$ values. The one neutron and two neutron transfer $Q$-values from projectile to target nuclei $Q_{1n}$ and $Q_{2n}$, respectively, are also listed while neutron transfers from target to projectile nuclei are provided within parentheses.}
\begin{tabular}{|c|c|c|c|c|c|c|c|}
\hline
Reaction &$Q$ (fusion)&$Q_{1n}$&$Q_{2n}$&$r_0$&$\sigma_1$&$\sigma_2$&$\sigma_0$ \\
& [MeV]& [MeV]& [MeV] & [fm]& [MeV] & [MeV] & [MeV]         \\  
\hline

$^{12}$C+$^{24}$Mg&16.298&-11.391 (-3.866) &-13.418 (-0.739) &1.285  &0.0000 &0.6417 &0.5025 \\   
$^{12}$C+$^{30}$Si&14.114&-12.134 (-5.663) &-16.051 (-5.960) &1.538  &0.0000 &0.3656 &1.0268  \\  
$^{12}$C+$^{198}$Pt&-13.955&-13.166 (-2.611) &-19.004 (-0.281) &1.377  &0.0000 &0.7992 &1.5557  \\ 
$^{16}$O+$^{18}$O&24.413&-11.709 (-3.910) &-17.323 (0.0) &1.506  &0.0116 &0.0119 &0.8588 \\
$^{28}$Si+$^{64}$Ni&-1.787&-11.082 (-1.185) &-15.441 (2.587) &1.021  &2.5738 &0.7806 &0.0000 \\
$^{58}$Ni+$^{58}$Ni&-65.855&-3.218 (-3.218) &-2.079 (-2.079) &1.104  &0.0000 &0.0000 &2.2750  \\ 
$^{64}$Ni+$^{64}$Ni&-48.797&-3.560 (-3.560) &-1.446 (-1.446) &1.108  &1.1823 &1.1823 &0.0000  \\
\hline
\end{tabular} 
\vspace{0.0cm}
\end{table*}

\subsection{ Fusion hindrance and neutron transfer $Q$-values }

    The one neutron and two neutron transfer $Q$-values from projectile to target nuclei $Q_{1n}$ and $Q_{2n}$, respectively, and those from target to projectile nuclei (except for $^{28}$Si+$^{64}$Ni) are all negative. The reactions studied in the present work, the transfer effects are, therefore, absent leading to fusion hindrance. Thence the width $\sigma_h$ of the distribution function is not enhanced due to any mixture of transfer and deformation effects. It is worthwhile to mention here that for reaction $^{32}$S + $^{110}$Pd for which neutron transfer $Q$-values are positive (and hence a mixture of transfer and deformation effects) leading to large $\sigma_h$ = 3.10 MeV which is significantly greater than $\sigma_h$ = 1.92 MeV for the reaction $^{36}$S + $^{110}$Pd for which neutron transfer $Q$-values are negative (and only pure deformation effect due to the deformed Pd) \cite{At14}. In general, for positive fusion $Q$-values, the fusion cross section should be more. But for the fusion reactions $^{12}$C+$^{24}$Mg, $^{12}$C+$^{30}$Si and $^{16}$O+$^{18}$O with positive $Q$-values, fusion hindrance remains owing to particularly small $\sigma_h$ values as is evident from Eqs.(3,4). Such an effect is due to the saturation properties of nuclear matter, which hinders density build up and prevents substantial overlap of light nuclei participating in reactions causing hindrance in quantum tunneling. This leads to rapid decrease in fusion cross section characterizing a major impact on the estimations of thermonuclear reaction rates which play a very significant role in stellar evolution studies.
        
\noindent
\section{ Summary and conclusion }
\label{section5}

    In the region of sub-barrier energies, the fusion reaction cross sections have been estimated spanning a broad energy range. In order to envision the conditions of overcoming the potential barrier in nuclear collisions and to have a systematic knowledge on the essential characteristics, {\it viz.} $h_0$ (mean barrier height), $\sigma_h$ (width) and $R_h$ (effective radius), of the interacting potential, a set of accurately measured excitation functions of fusion reactions has been studied for two colliding nuclei. A Gaussian distribution function for the barrier heights is assumed to derive a simple diffused-barrier formula. The values of the essential parameters $h_0$ (mean barrier height), $\sigma_h$ (width) and $R_h$ (effective radius) are determined using the method of least-square fit. In the fusion reactions studied here, the transfer effects are absent leading to fusion hindrance. The widths of the barrier distribution are not enhanced due to any mixture of transfer and deformation effects. Even for positive fusion $Q$-values, the fusion hindrance remains because of small $\sigma_h$. This effect may be attributed to the saturation properties of nuclear matter, which prevents substantial overlap of light nuclei participating in the fusion reactions causing hindrance in quantum tunneling. 
  
    The present formula of cross section for fusion reactions can be used to calculate the cross sections for surmounting the barrier in collisions of moderately heavy systems for a given projectile-target combination. For calculating the production cross sections of superheavy nuclei, the prediction of the capture excitation functions or sticking can be used in the sticking-diffusion-survival model \cite{An93} as one of three basic ingredients. The reasonably good fit to the experimental data provided by the theory described above implies two principal facts that for the investigated set of nucleus-nucleus systems almost all the events leading to capture ultimately proceeds to fusion and the idea of the Gaussian distribution of barrier provides excellent description of cross sections for fusion reactions in the domain of the sub-barrier energies. Although the single-Gaussian parametrization for barrier distribution is reasonably successful in providing a good description of fusion process in general, neither the formula derived for fusion cross-section nor the method using the barrier distribution can be put to use for all fusing systems. One can visualize from Eq.(3) that the excitation function is all the time a monotonically rising function of energy. This puts a limitation on Eqs.(1,2) which can not be used for describing fusion reactions at higher energies when incomplete fusion as well as deep-inelastic scattering can cause a lowering of the fusion cross section. Similar limitation can arise for lighter systems also when excitation functions possess oscillations and resonance structures. Possibility of better compliance to measured data may be explored by opting a more intricate formula for barrier distribution, which, however, will bring in more additional adjustable parameters than just three used in the present work. Such improvements for the barrier distribution can be realized through distributions having distinctive widths on the lower or higher energy sides, multi-component distributions or a modification of the exponent appearing in distribution represented by a Gaussian form. 
   
\begin{acknowledgements}

    One of the authors (DNB) acknowledges support from Science and Engineering Research Board, Department of Science and Technology, Government of India, through Grant No. CRG/2021/007333.

\end{acknowledgements}

\end{document}